\documentclass[onecolum,aps,pre,reprint,showpacs,superscriptaddress]{revtex4-1}

\usepackage{CJK}
\usepackage{amsfonts}
\usepackage{amsmath}
\usepackage{tipa}
\usepackage{amssymb}
\usepackage{epsfig}
\usepackage{bm,color}
\begin{document}

\title{Could the Earth's surface Ultraviolet irradiance be blamed for the global warming?(II)\\
        \textit{------Ozone layer depth reconstruction via HEWV effect}}

\author{Jilong Chen}
\email{niba180634@gmail.com}
\affiliation{Department of Astronomy, Beijing Normal University,Beijing 100875, China}

\author{Juan Zhao}
\email{zj@bnu.edu.cn}
\affiliation{Department of Astronomy, Beijing Normal University,Beijing 100875, China}

\author{Yujun Zheng}
\email{yzheng@sdu.edu.cn}
\affiliation{School of Physics, Shandong University, Jinan 250100, China}

\begin{abstract}
It is suggested by Chen {\it et al.} that the Earth's surface Ultraviolet
irradiance ($280-400$ nm) could influence the Earth's surface temperature
variation by ``Highly Excited Water Vapor" (HEWV) effect.
In this manuscript, we reconstruct the developing history of the ozone layer depth
variation from 1860 to 2011 based on the HEWV effect. It is shown that
the reconstructed ozone layer depth variation correlates with the observational
variation from 1958 to 2005 very well ($R=0.8422$, $P>99.9\%$).
From this reconstruction, we may limit the spectra band of the surface
Ultraviolet irradiance referred in HEWV effect to Ultraviolet B ($280-320$ nm).

\end{abstract}


\maketitle

\section{introduction}
\label{sec:int}

\begin{figure} [h]
 \centering
  \includegraphics[width=0.8\hsize]{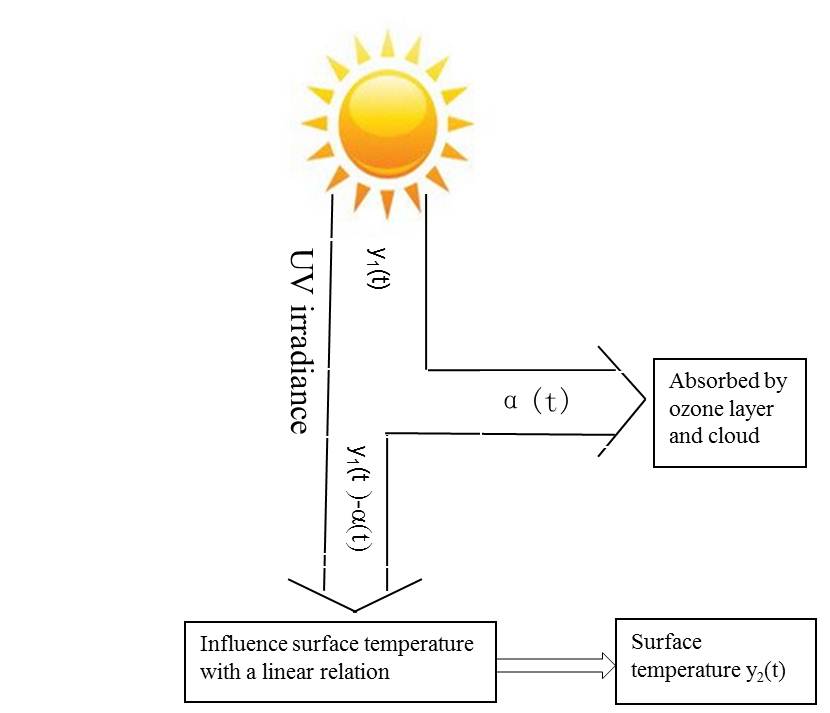}
  \caption{The schematic diagram of physical relation between
  solar UV irradiance and surface temperature}
\label{Fig:phys-rel}
\end{figure}

It is suggested that the surface solar
Ultraviolet (UV) irradiance ($280-400$ nm, similarly hereinafter)
may have the ability to influence the Earth's surface temperature by
``Highly Excited Water Vapor" (HEWV) effect~\cite{chen}.
Based on the HEWV effect, it is schematically shown in Fig.~\ref{Fig:phys-rel}
that the physical way of solar UV irradiance influences
the Earth's surface temperature: Part of the
UV irradiance emitted by the Sun is absorbed by the Earth's ozone layer and
cloud, while the rest reaches to the lower troposphere and influences the surface
temperature via HEWV effect. In this manuscript, we suggest a model to obtain the
UV irradiance absorbed by ozone layer and the cloud by employing the Earth's surface
temperature data and solar UV irradiance data at the top of the Earth's
atmosphere. The UV irradiance absorbed by ozone layer and the cloud could
describe the variation of ozone layer or cloud.

\section{the model}
\label{sec:theory}

\begin{figure} [h] 
  \centering
  \includegraphics[width=0.9\hsize]{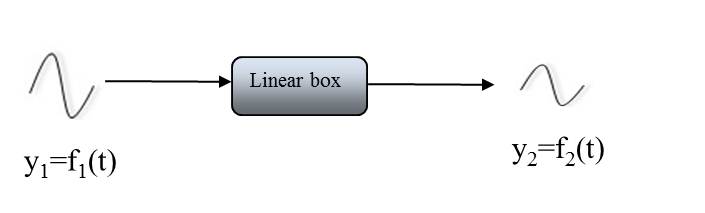}
  \caption{Diagrammatic sketch of the model system.
  Signal $y_1$ changes to $y_2$ after getting through the linear system}
  \label{Fig:model}
\end{figure}

In our model, we assume a signal $y_1=f_1(t)$ goes through a linear system box,
which is in a steady state. It is schematically shown in Fig.~\ref{Fig:model}.
Then, we can obtain a new signal $y_2=f_2(t)$,
and $y_1$ has a linear relation with $y_2$, namely
\begin{equation}
y_1=Ky_2+b.
\label{eq:model}
\end{equation}

Based on HEWV effect, we suggest the surface UV irradiance influences
the surface temperature linearly.
Equation~(\ref{eq:model}) can be employed to describe the relation between surface
temperature and surface UV irradiance in spectra band $280-400$ nm.
In our model, the Sun emits $y_1(t)$ solar UV irradiance in spectra band 280-400 nm to
the top of the Earth's atmosphere at the time $t$, and $\alpha(t)$
is absorbed by ozone layer and cloud. The rest part $y_1(t)-\alpha(t)$ goes to
the lower troposphere and influences the surface temperature $y_2(t)$ in
a linear relation:
\begin{equation}
y_1(t)-\alpha(t)=Ky_2(t)+b ,
\label{eq:alp}
\end{equation}
or, it can be rewritten as follow:
\begin{equation}
\alpha(t)=y_1(t)-Ky_2(t)-b.
\label{eq:alpha}
\end{equation}

The solar UV irradiance data in spectra band $280-400$ nm $y_1(t)$ and
surface temperature data $y_2(t)$ can be obtained.
The amounts  absorbed by ozone layer and cloud $\alpha(t)$
can be obtained from Eq.~(\ref{eq:alpha}) for given $K$ and $b$,
which can describe the variation of the ozone layer or cloud.

An appropriate way to obtain $K$ and $b$ is to find the years when
the ozone layer and cloud did not change (i.e. $ \rm \alpha(t)=\alpha_0=\rm{constant}$).
If this happened, Eq.~(\ref{eq:alp}) can be rewritten as
\begin{equation}
y_1(t)=Ky_2(t)+C ,
\label{eq:y-c}
\end{equation}
where $C=b+\alpha_0$. Eq.~(\ref{eq:y-c}) can be employed to obtain the coefficients $K$
and $C$ using the data of $y_1$ and $y_2$ during these years.
Once we know the values of $K$ and $C$,
the UV irradiance anomaly absorbed by the ozone layer and the cloud, $\Delta\alpha(t)$,
can be obtained as follows
\begin{equation}
\Delta\alpha(t)=y_1(t)-Ky_2(t)-C ,
\label{eq:moment}
\end{equation}
where $\Delta\alpha(t)=\alpha(t)-\alpha_0$.
It is clearly that $\rm\Delta\alpha$ can also describe
the variation of ozone layer or cloud.

However, we could only find the years when the cloud
and ozone layer are keeping the same roughly. This means
we could just obtain $ \rm \alpha(t)\approx\alpha_0$.
In this case we can search for the years when the $y_1(t)$ correlated with $y_2(t)$ well,
and employ the linear fitting to obtain the values of $K$ and $C$.

Based on Eq.~(\ref{eq:moment}),  HEWV effect can be shown via comparing $\Delta\alpha$ with
the observed ozone layer depth data or cloud data.

\section{Data}

In this manuscript, the detail data of the surface temperature and
solar UV irradiance can be taken from the public data source, they are listed as below:

\subsection{Surface temperature (ST)}
Global yearly mean surface temperature anomaly from 1860 to 2011 is
obtained from HadCRUT3 in Climate Research Unit (CRU) \cite{JGRD}.

\subsection{Solar UV irradiance}
Solar Spectra Irradiance (SSI) variation in spectra band $280-400$ nm
from 1860 to 2005 is reconstructed by Krivova {\it et al.} \cite{krivova2010}.

\subsection{Total Solar irradiance (TSI)}
Also, the Total Solar Irradiance (TSI) can also be employed to reconstruct $ \Delta\alpha$,
because the variation of the TSI can describe the property of the SSI in
spectra band $280-400$ nm very well. We test the relation between reconstructed SSI
and TSI from 1860 to 2005 and find they correlate with each other very well ($R=0.8402$, $P>99.9$\%).
Moreover, we can obtain satellite observational data after 1979, which could make the
value closer to the real TSI variation and may reconstruct the $ \Delta\alpha$ more accurate.

The TSI data from 1850 to 1978 was reconstructed by Judith Lean \cite{lean1995}.
Satellite data from 1979 to 2003 is from the VIRGO Experiment on the
cooperative ESA/NASA Mission SoHO \cite{virgo}.
Satellite data from 2002 to 2011 is from SORCE TIM Total Solar Irradiance \cite{sorce}.

\subsection{Ozone layer depth}
The annual average of global ($65 \rm{^o S}-65 \rm{^o N}$) column total
ozone variation from 1958 to 2005 was compiled by
Johnston \cite{Robert}, which was synthesized the
Ground-based observations data and the satellite observations data.
The Ground-based observations data from 1958 to 1977 was analyzed by Angell
and Korshover in Carbon Dioxide Information Analysis Center (CDIAC) \cite{Angell}.
The 1978 to 1992 satellite data was presented by Total Ozone Mapping Spectrometer (TOMS),
NASA, measured by  Nimbus-7 satellite \cite{toms}.
The ozone layer data from 1993 to 1994 was obtained from  Meteor-3
satellite by TOMS, NASA \cite{meteor3}. The 1995 data was analyzed by Weber \cite{weber}.
The ozone layer data from 1996 to 2005 was presented by TOMS \cite{eptoms}.

\section{Results}

\subsection{Reconstruct ozone layer depth by employing SSI and ST}

\begin{figure} 
  \centering
  \includegraphics[width=0.9\hsize]{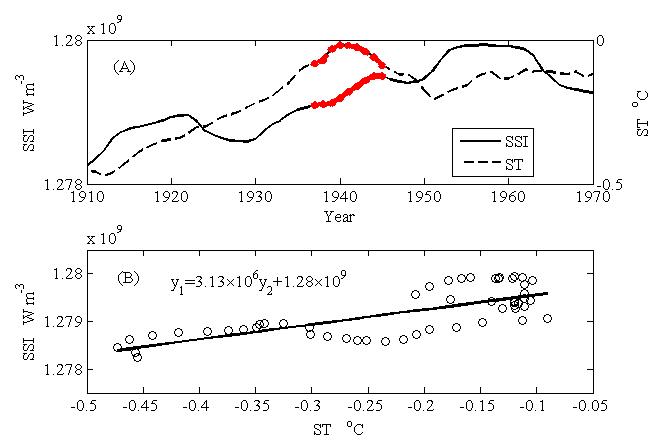}
  \caption{(A) Comparison between Solar spectra irradiance (SSI) in
  spectra band $280-400$ nm and surface temperature (ST) from 1910 to 1970
  (both data are in $11$-yr moving average), data from 1937 to 1945 being
  deducted (in red color); (B) Linear fit of the two series.}
  \label{Fig:ssifit}
\end{figure}

Figure~\ref{Fig:ssifit}(A) shows the comparison of ST ($y_2$) and SSI ($y_1$) variation
from 1910 to 1970 in 11-yr moving average. Good correlation is found
between them ($R=0.6750$, $P>99.9$\%), so we suppose ozone layer or cloud did not
change much in this period (we deduct the time series during 1937 and 1945,
in the global war II, when there were not enough observational data \cite{kap1998})
and employ them to calculate the parameters $K$ and $C$ using linear fitting method.
The linear fitting result is shown in Fig.~\ref{Fig:ssifit}(B).
The parameters: $K=3.37\times10^{6}$ and $C=1.28\times10^{9}$.

\begin{figure} 
  \centering
  \includegraphics[width=0.9\hsize]{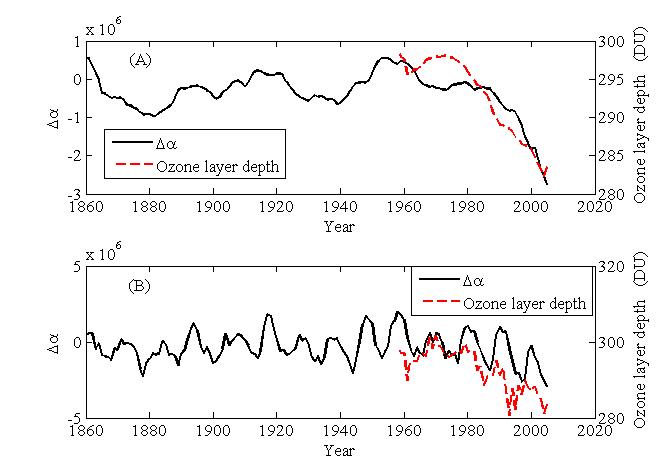}
  \caption{Variation of $\rm\Delta\alpha$ from 1860 to 2005 (by employing SSI).
  (A): The 11-yr moving average variation of $\rm\Delta\alpha$, compared with
  the observed 11-yr moving average ozone layer depth variation from 1958 to 2005.
  (B): Yearly variation of $\rm\Delta\alpha$, compared with the yearly variation
  of observational ozone layer depth from 1958 to 2005}
  \label{Fig:dalpha_ssi}
\end{figure}

We can use the parameters $K$ and $C$ to calculate the variation
of  $\rm\Delta\alpha$ based on Eq.~(\ref{eq:moment}).
The result of $\rm\Delta\alpha$ is shown in Fig.~\ref{Fig:dalpha_ssi}.
Fig.~\ref{Fig:dalpha_ssi}(A) shows the 11-yr moving average variation
from 1860 to 2005, compared with observed 11-yr moving average ozone layer
depth variation from 1958 to 2005. We find a good correlation between
them ($R=0.8965$, $P>99.9$\%). Fig.~\ref{Fig:dalpha_ssi}(B) shows the
yearly variation of $\rm\Delta\alpha$, compared with the yearly ozone
layer depth observed variation. Good correlation between them is also found: $R=0.6323$, $P>99.9$\%.

\begin{figure} 
  \centering
  \includegraphics[width=0.9\hsize]{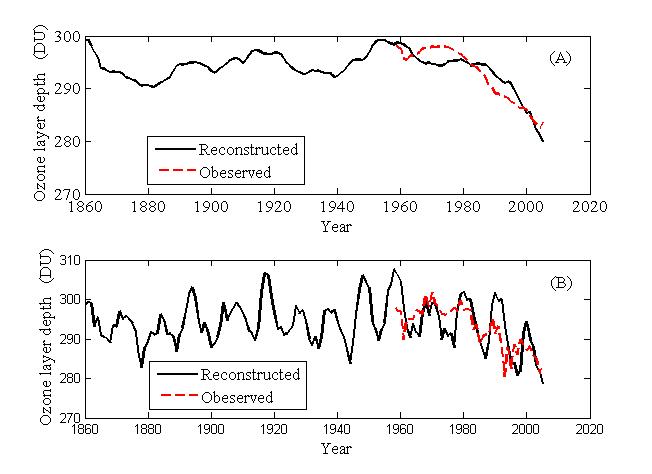}
  \caption{Reconstructed ozone layer depth variation from 1860 to 2005 (by employing SSI data),
  compared with the observational data. (A): In 11-yr moving average. (B): Yearly variation}
  \label{Fig:re_ssi}
\end{figure}

Based on the good correlation above, we can reconstruct the ozone layer depth
variation from 1860 to 2005. Firstly, we fit the linear relation
between $\rm\Delta\alpha$ and ozone layer depth by applying $\rm\Delta\alpha$ and
observational ozone layer depth data from 1958 to 2005, and it can be written as follows
\begin{equation}
y_o=5.83 \times 10^{-6} \Delta\alpha+295.99,
\label{eq:re_ssi}
\end{equation}
where $y_o$ is the ozone layer depth. We expand the relation described by
Eq.~(\ref{eq:re_ssi}) to the whole $\rm\Delta\alpha$  and get the reconstructed
ozone layer variation from 1860 to 2005. This is shown in Fig.~\ref{Fig:re_ssi}.

\subsection{Reconstruct ozone layer depth by employing TSI and ST}

\begin{figure} 
  \centering
  \includegraphics[width=0.9\hsize]{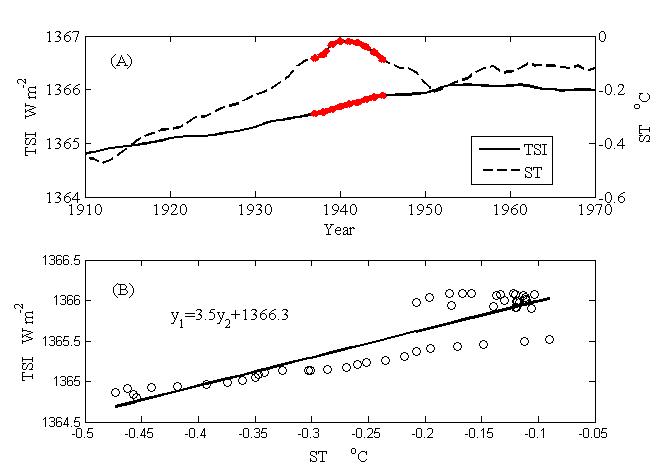}
  \caption{Comparison between Total Solar irradiance (TSI) and surface
  temperature (ST) from 1910 to 1970, data from 1937 to 1945 being
  deducted (in red color); (B) Linear fitting of the two series.}
  \label{Fig:dalpha_tsi}
\end{figure}

The TSI $(y_1)$ and ST $(y_2)$ data from 1910 to 1970 are applied
(time series during 1937 and 1945 is deducted)
(See Fig.~\ref{Fig:dalpha_tsi}(A)).
We employ linear fitting and get $K=3.5$ and $C= 1366.3$
(See Fig.~\ref{Fig:dalpha_tsi}(B)).

\begin{figure} 
  \centering
  \includegraphics[width=0.9\hsize]{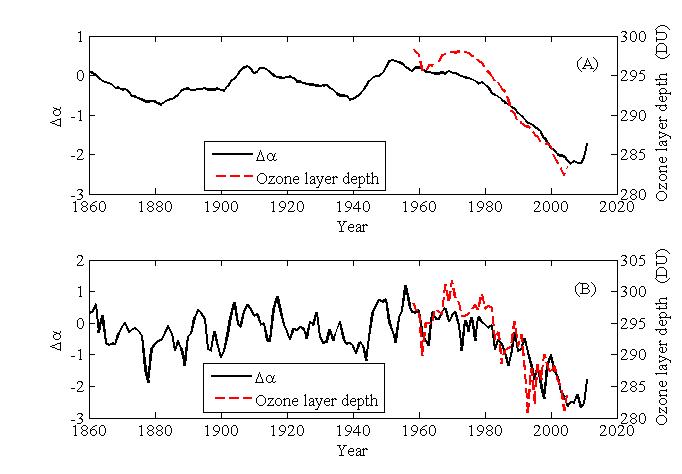}
  \caption{Variation of $\rm\Delta\alpha$ from 1860 to 2011 (by employing TSI).
  (A): The 11-yr moving average variation of $\rm\Delta\alpha$,
  compared with the observed 11-yr moving average ozone layer depth variation from 1958 to 2005.
  (B): Yearly variation of $\rm\Delta\alpha$, compared with the yearly variation of
  observational ozone layer depth from 1958 to 2005}
  \label{Fig:re_tsi}
\end{figure}

$K$ and $C$ are employed to calculate the variation
of  $\rm\Delta\alpha$ based on Eq.(~\ref{eq:moment}).
The result of $\rm\Delta\alpha$ is shown in Fig.~\ref{Fig:re_tsi}.
Fig.~\ref{Fig:re_tsi}(A) shows the 11-yr moving average variation
from 1860 to 2011, compared with observed 11-yr moving average ozone
layer depth variation from 1958 to 2005.
We find a good correlation between them ($R=0.9822$, $P>99.9$\%).
Fig.~\ref{Fig:re_tsi}(B) shows the yearly variation of $\rm\Delta\alpha$,
compared with the yearly ozone layer depth observed variation.
Good correlation between them is also found ($R=0.8422$, $P>99.9$\%).
This result of $\rm\Delta\alpha$ is much better than that in SSI occasion.

\begin{figure} 
  \centering
  \includegraphics[width=0.9\hsize]{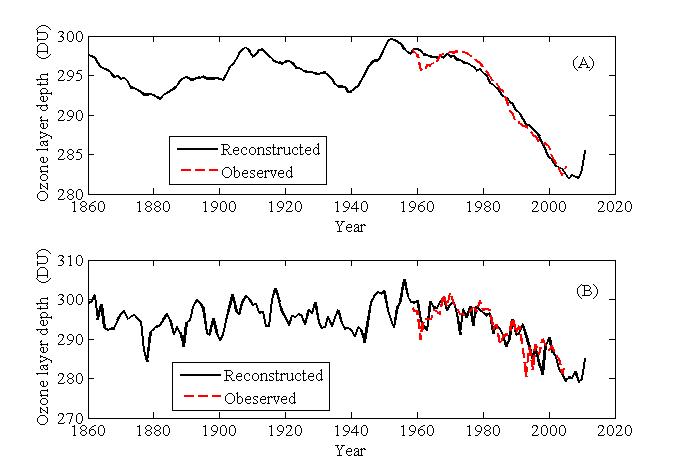}
  \caption{Reconstructed ozone layer variation (using TSI) compared with
  the observed ozone layer variation.
  (A): In 11-yr moving average; (B) Yearly variation}
  \label{Fig:tsi}
\end{figure}

We could also reconstruct the ozone layer depth variation from 1860 to 2011
based on the linear fitting relation between $\rm\Delta\alpha$ and
observed ozone layer depth from 1958 to 2005:
\begin{equation}
y_o=6.72\Delta\alpha+297.00.
\label{eq:re_tsi}
\end{equation}
We expand the relation described by Eq.~(\ref{eq:re_tsi}) to
the whole $\rm\Delta\alpha$  and get the reconstructed ozone layer
variation from 1860 to 2011 as shown in Fig.~\ref{Fig:tsi}.

\section{discussion}

We could see from Fig.~\ref{Fig:tsi} that the reconstructed ozone layer depth
data describes the property of the ozone layer depth variation very well,
but there are still some differences, such as the opposite variation during 1993 and 2000.
The difference may led by the volcano eruption in 1991 and El Ni\~{n}o and
La Ni\~{n}o in 1997-1999. Researchers have pointed the cooling effect of
the volcanic eruption to the surface temperature, such as Robock and Mao
shows that for two years following great volcanic eruptions,
the surface of the Earth could be cooled significantly by $0.1\rm{^oC} \sim 0.2\rm{^oC}$ in
the global mean scale \cite{robock1995}. In 1991, the volcanic eruption of
Mount Pinatubo occurred, cooling the Earth's surface in the following $2$ years,
which will lead to a lower value of the reconstructed surface UV irradiance,
then a higher UV irradiance absorbed by ozone layer  in the reconstruction,
and eventually cause a higher reconstructed ozone layer depth near 1993.
This suggestion could also interpret the higher reconstructed ozone layer depth data in 1985,
when the volcanic eruption in 1983. El Ni\~{n}o could lead a warmer phenomenon in
the Earth's surface \cite{robock1995}. The El Ni\~{n}o appearance in 1997-1998 could
lead a warmer surface temperature, which will lead to a lower value of
the reconstructed ozone layer depth variation near 1998.
We can see the same phenomenon in the year of 1987-1988, when the El Ni\~{n}o occurred.

\begin{figure} [h]
  \centering
  \includegraphics[width=0.8\hsize]{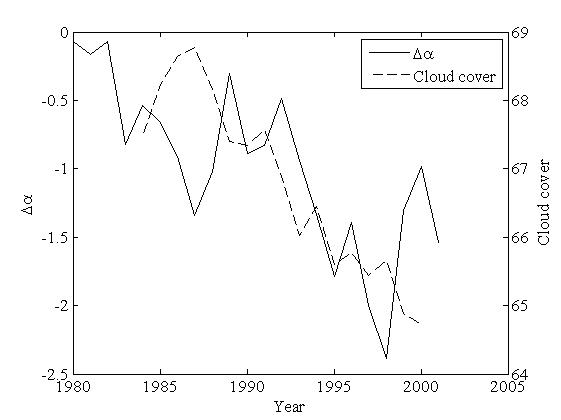}
  \caption{Comparison of the cloud cover variation (Unit: \%) and $\rm\Delta\alpha$}
  \label{Fig:cloud}
\end{figure}

The signal of the cloud cover doesn't be recognized obviously.
Cloud cover is compared with the $\rm\Delta\alpha$ in Fig.~\ref{Fig:cloud}.
The global mean cloud coverage data is obtained from International
Satellite cloud climatology Project (ISCCP), NASA \cite{isccp}.
They don't show an apparent correlation  ($R=0.4995$, $P>95$\%),
as obviously as the $\rm\Delta\alpha$ with ozone layer depth ($R=0.8422$, $P>99.9$\%).
The reason why the cloud cover signal doesn't been recognized obviously
need to do more researches. But based on the results,
we may suggest that the ozone layer plays more important role in the HEWV effect.
The spectra band absorbed by ozone layer is UVB ($280-320$ nm),
and this result may limit the spectra that works in HEWV effect to UVB irradiance.
More deeper researches should be done before we get a more clear conclusion.

\begin{acknowledgements}
This work is supported by National Natural Science Foundation of
China (Grant Nos. 11203004, 10978007, 11374191 and 91021009).
\end{acknowledgements}



\begin{thebibliography}{15}%
\makeatletter
\providecommand \@ifxundefined [1]{%
 \@ifx{#1\undefined}
}%
\providecommand \@ifnum [1]{%
 \ifnum #1\expandafter \@firstoftwo
 \else \expandafter \@secondoftwo
 \fi
}%
\providecommand \@ifx [1]{%
 \ifx #1\expandafter \@firstoftwo
 \else \expandafter \@secondoftwo
 \fi
}%
\providecommand \natexlab [1]{#1}%
\providecommand \enquote  [1]{``#1''}%
\providecommand \bibnamefont  [1]{#1}%
\providecommand \bibfnamefont [1]{#1}%
\providecommand \citenamefont [1]{#1}%
\providecommand \href@noop [0]{\@secondoftwo}%
\providecommand \href [0]{\begingroup \@sanitize@url \@href}%
\providecommand \@href[1]{\@@startlink{#1}\@@href}%
\providecommand \@@href[1]{\endgroup#1\@@endlink}%
\providecommand \@sanitize@url [0]{\catcode `\\12\catcode `\$12\catcode
  `\&12\catcode `\#12\catcode `\^12\catcode `\_12\catcode `\%12\relax}%
\providecommand \@@startlink[1]{}%
\providecommand \@@endlink[0]{}%
\providecommand \url  [0]{\begingroup\@sanitize@url \@url }%
\providecommand \@url [1]{\endgroup\@href {#1}{\urlprefix }}%
\providecommand \urlprefix  [0]{URL }%
\providecommand \Eprint [0]{\href }%
\providecommand \doibase [0]{http://dx.doi.org/}%
\providecommand \selectlanguage [0]{\@gobble}%
\providecommand \bibinfo  [0]{\@secondoftwo}%
\providecommand \bibfield  [0]{\@secondoftwo}%
\providecommand \translation [1]{[#1]}%
\providecommand \BibitemOpen [0]{}%
\providecommand \bibitemStop [0]{}%
\providecommand \bibitemNoStop [0]{.\EOS\space}%
\providecommand \EOS [0]{\spacefactor3000\relax}%
\providecommand \BibitemShut  [1]{\csname bibitem#1\endcsname}%
\let\auto@bib@innerbib\@empty
\bibitem [{\citenamefont {Chen}\ \emph {et~al.}(2014)\citenamefont {Chen},
  \citenamefont {Sun}, \citenamefont {Zhao},\ and\ \citenamefont
  {Zheng}}]{chen}%
  \BibitemOpen
  \bibfield  {author} {\bibinfo {author} {\bibfnamefont {J.}~\bibnamefont
  {Chen}}, \bibinfo {author} {\bibfnamefont {Z.}~\bibnamefont {Sun}}, \bibinfo
  {author} {\bibfnamefont {J.}~\bibnamefont {Zhao}}, \ and\ \bibinfo {author}
  {\bibfnamefont {Y.}~\bibnamefont {Zheng}},\ }\href@noop {} {\  (\bibinfo
  {year} {2014})},\ \Eprint {http://arxiv.org/abs/arXiv:1411.6511v1}
  {arXiv:1411.6511v1} \BibitemShut {NoStop}%
\bibitem [{\citenamefont {Brohan}\ \emph {et~al.}(2006)\citenamefont {Brohan},
  \citenamefont {Kennedy}, \citenamefont {Harris}, \citenamefont {Tett},\ and\
  \citenamefont {Jones}}]{JGRD}%
  \BibitemOpen
  \bibfield  {author} {\bibinfo {author} {\bibfnamefont {P.}~\bibnamefont
  {Brohan}}, \bibinfo {author} {\bibfnamefont {J.~J.}\ \bibnamefont {Kennedy}},
  \bibinfo {author} {\bibfnamefont {I.}~\bibnamefont {Harris}}, \bibinfo
  {author} {\bibfnamefont {S.~F.~B.}\ \bibnamefont {Tett}}, \ and\ \bibinfo
  {author} {\bibfnamefont {P.~D.}\ \bibnamefont {Jones}},\ }\href {\doibase
  10.1029/2005JD006548} {\bibfield  {journal} {\bibinfo  {journal} {Journal of
  Geophysical Research: Atmospheres}\ }\textbf {\bibinfo {volume} {111}},\
  \bibinfo {pages} {n/a} (\bibinfo {year} {2006})}\BibitemShut {NoStop}%
\bibitem [{\citenamefont {Krivova}\ \emph {et~al.}(2010)\citenamefont
  {Krivova}, \citenamefont {Vieira},\ and\ \citenamefont
  {Solanki}}]{krivova2010}%
  \BibitemOpen
  \bibfield  {author} {\bibinfo {author} {\bibfnamefont {N.}~\bibnamefont
  {Krivova}}, \bibinfo {author} {\bibfnamefont {L.}~\bibnamefont {Vieira}}, \
  and\ \bibinfo {author} {\bibfnamefont {S.}~\bibnamefont {Solanki}},\
  }\href@noop {} {\bibfield  {journal} {\bibinfo  {journal} {Journal of
  Geophysical Research: Space Physics (1978--2012)}\ }\textbf {\bibinfo
  {volume} {115}} (\bibinfo {year} {2010})}\BibitemShut {NoStop}%
\bibitem [{\citenamefont {Lean}\ \emph {et~al.}(1995)\citenamefont {Lean},
  \citenamefont {Beer},\ and\ \citenamefont {Bradley}}]{lean1995}%
  \BibitemOpen
  \bibfield  {author} {\bibinfo {author} {\bibfnamefont {J.}~\bibnamefont
  {Lean}}, \bibinfo {author} {\bibfnamefont {J.}~\bibnamefont {Beer}}, \ and\
  \bibinfo {author} {\bibfnamefont {R.}~\bibnamefont {Bradley}},\ }\href@noop
  {} {\bibfield  {journal} {\bibinfo  {journal} {Geophysical Research Letters}\
  }\textbf {\bibinfo {volume} {22}},\ \bibinfo {pages} {3195} (\bibinfo {year}
  {1995})}\BibitemShut {NoStop}%
\bibitem [{\citenamefont {Experiment}(2003)}]{virgo}%
  \BibitemOpen
  \bibfield  {author} {\bibinfo {author} {\bibfnamefont {VIRGO}~\bibnamefont
  {Experiment}},\ }\href@noop {} {\enquote {\bibinfo {title} {Soho data},}\
  }\bibinfo {howpublished} {\url{http://sohowww.nascom.nasa.gov/data/}}
  (\bibinfo {year} {2003})\BibitemShut {NoStop}%
\bibitem [{\citenamefont {NASA}(2013)}]{sorce}%
  \BibitemOpen
  \bibfield  {author} {\bibinfo {author} {\bibfnamefont {SORCE}~\bibnamefont
  {NASA}},\ }\href@noop {} {\enquote {\bibinfo {title} {Sorce $tim$ total solar
  irradiance},}\ }\bibinfo {howpublished}
  {\url{http://lasp.colorado.edu/home/sorce/data/tsi-data/#plots}} (\bibinfo
  {year} {2013})\BibitemShut {NoStop}%
\bibitem [{\citenamefont {Johnston}(2006)}]{Robert}%
  \BibitemOpen
  \bibfield  {author} {\bibinfo {author} {\bibfnamefont {W.~R.}\ \bibnamefont
  {Johnston}},\ }\href@noop {} {\enquote {\bibinfo {title} {Historical data
  relating to the ozone layer},}\ }\bibinfo {howpublished}
  {\url{http://www.johnstonsarchive.net/environment/o3cltable.html}} (\bibinfo
  {year} {2006})\BibitemShut {NoStop}%
\bibitem [{\citenamefont {Angell}\ and\ \citenamefont
  {Korshover}(1989)}]{Angell}%
  \BibitemOpen
  \bibfield  {author} {\bibinfo {author} {\bibfnamefont {J.~K.}\ \bibnamefont
  {Angell}}\ and\ \bibinfo {author} {\bibfnamefont {J.}~\bibnamefont
  {Korshover}},\ }\href@noop {} {\enquote {\bibinfo {title} {Annual and
  seasonal global variation in total ozone and layer-mean ozone, 1958-1987},}\
  }\bibinfo {howpublished} {\url{http://cdiac.ornl.gov/ftp/ndp023/ndp023.txt}}
  (\bibinfo {year} {1989})\BibitemShut {NoStop}%
\bibitem [{\citenamefont {TOMS}(2006{\natexlab{a}})}]{toms}%
  \BibitemOpen
  \bibfield  {author} {\bibinfo {author} {\bibnamefont {TOMS}},\ }\href@noop {}
  {\enquote {\bibinfo {title} {Nimbus 7 toms zonal means},}\ }\bibinfo
  {howpublished}
  {\url{ftp://toms.gsfc.nasa.gov/pub/nimbus7/data/zonal_means/ozone/ZM_month_ozone_N7.txt}}
  (\bibinfo {year} {2006}{\natexlab{a}})\BibitemShut {NoStop}%
\bibitem [{\citenamefont {TOMS}(2006{\natexlab{b}})}]{meteor3}%
  \BibitemOpen
  \bibfield  {author} {\bibinfo {author} {\bibnamefont {TOMS}},\ }\href@noop {}
  {\enquote {\bibinfo {title} {Meteor-3 toms zonal means},}\ }\bibinfo
  {howpublished}
  {\url{ftp://toms.gsfc.nasa.gov/pub/meteor3/data/zonal_means/ozone/ZM_month_ozone_M3.txt}}
  (\bibinfo {year} {2006}{\natexlab{b}})\BibitemShut {NoStop}%
\bibitem [{\citenamefont {Weber}()}]{weber}%
  \BibitemOpen
  \bibfield  {author} {\bibinfo {author} {\bibfnamefont {M.}~\bibnamefont
  {Weber}},\ }\href@noop {} {\enquote {\bibinfo {title} {Monthly and zonal mean
  gome wfdoas total ozone},}\ }\bibinfo {howpublished}
  {\url{http://www.iup.uni-bremen.de/gome/wfdoas/overpass/}}\BibitemShut
  {NoStop}%
\bibitem [{\citenamefont {TOMS}()}]{eptoms}%
  \BibitemOpen
  \bibfield  {author} {\bibinfo {author} {\bibnamefont {TOMS}},\ }\href@noop {}
  {\enquote {\bibinfo {title} {Earth probe zonal mean of total ozone},}\
  }\bibinfo {howpublished}
  {\url{ftp://toms.gsfc.nasa.gov/pub/eptoms/data/zonal_means/ozone/ZM_month_ozone_ept.txt}}\BibitemShut
  {NoStop}%
\bibitem [{\citenamefont {Kaplan}\ \emph {et~al.}(1998)\citenamefont {Kaplan},
  \citenamefont {Cane}, \citenamefont {Kushnir}, \citenamefont {Clement},
  \citenamefont {Blumenthal},\ and\ \citenamefont {Rajagopalan}}]{kap1998}%
  \BibitemOpen
  \bibfield  {author} {\bibinfo {author} {\bibfnamefont {A.}~\bibnamefont
  {Kaplan}}, \bibinfo {author} {\bibfnamefont {M.~A.}\ \bibnamefont {Cane}},
  \bibinfo {author} {\bibfnamefont {Y.}~\bibnamefont {Kushnir}}, \bibinfo
  {author} {\bibfnamefont {A.~C.}\ \bibnamefont {Clement}}, \bibinfo {author}
  {\bibfnamefont {M.~B.}\ \bibnamefont {Blumenthal}}, \ and\ \bibinfo {author}
  {\bibfnamefont {B.}~\bibnamefont {Rajagopalan}},\ }\href@noop {} {\bibfield
  {journal} {\bibinfo  {journal} {Journal of Geophysical Research: Oceans}\
  }\textbf {\bibinfo {volume} {103}},\ \bibinfo {pages} {18567} (\bibinfo
  {year} {1998})}\BibitemShut {NoStop}%
\bibitem [{\citenamefont {Robock}\ and\ \citenamefont
  {Mao}(1995)}]{robock1995}%
  \BibitemOpen
  \bibfield  {author} {\bibinfo {author} {\bibfnamefont {A.}~\bibnamefont
  {Robock}}\ and\ \bibinfo {author} {\bibfnamefont {J.}~\bibnamefont {Mao}},\
  }\href@noop {} {\bibfield  {journal} {\bibinfo  {journal} {Journal of
  Climate}\ }\textbf {\bibinfo {volume} {8}},\ \bibinfo {pages} {1086}
  (\bibinfo {year} {1995})}\BibitemShut {NoStop}%
\bibitem [{\citenamefont {ISCCP}\ and\ \citenamefont {NASA}()}]{isccp}%
  \BibitemOpen
  \bibfield  {author} {\bibinfo {author} {\bibnamefont {ISCCP,}}\ \bibinfo
  {author} {\bibnamefont {NASA}},\ }\href@noop {} {\enquote {\bibinfo {title}
  {Nasa isccp d$_2$ all mean cloud amount data files},}\ }\bibinfo
  {howpublished}
  {\url{http://iridl.ldeo.columbia.edu/SOURCES/.NASA/.ISCCP/.D2/.all/.mean_cloud/.amount/datafiles.html}}\BibitemShut
  {NoStop}%
\end{thebibliography}
%

\end{document}